\documentclass[twocolumn,secnumarabic,amssymb, nobibnotes, aps, prl]{revtex4-1} 

\usepackage{graphicx}
\usepackage{amssymb}

\usepackage{color}
\usepackage{xcolor}

\definecolor{green}{rgb}{0.0, 0.666, 0.0}

\definecolor{darkgreen}{rgb}{0.55, 0.71, 0.0}

\definecolor{lightviolet}{rgb}{0.79, 0.28, 0.96}

\definecolor{navy}{rgb}{0.05, 0.40, 0.8}

\definecolor{purple}{rgb}{0.62, 0.0, 0.666}

\begin{document}

\title{Shape-free theory for the self-assembly kinetics in macromolecular systems}%

\author{L. F. Trugilho}%
\author{L. G. Rizzi}%
\affiliation{Departamento\,de\,F\'isica, Universidade\,Federal\,de\,Vi\c{c}osa\,(UFV), \\
Av.\,P.\,H.\,Rolfs,~s/n,~36570-900,~Vi\c{c}osa,~Brazil.}

\begin{abstract}
Self-assembly kinetics is usually described by approaches which assume that the shape of the aggregates
has a definite form ({\it e.g.}, spherical, cylindrical, cubic, etc), however that is unlikely to be the case in many
finite-sized macromolecular and colloidal systems.
	Here we consider a simple
aggregation model which displays a first-order phase transition in order to illustrate a rate theory based on
microcanonical analysis that allows one to obtain a shape-free description of its self-assembly kinetics.
	Stochastic simulations are performed to validate our approach and demonstrate how
the equilibrium thermostatistical properties of the system can be related to the
temperature-dependent rate constants.
	As a model-independent kinetic approach, it may provide experimentalists a reliable method to reconstruct free-energy profiles and microcanonical entropies from kinetic data.
\end{abstract}

\maketitle


	Self-assembly kinetics is of particular interest to a myriad of scientific areas, 
ranging from climate and materials sciences to soft matter physics and biology~\cite{hagan2021revmodphys}.
	Nucleation, in particular, occurs when the system present 
interactions that lead to 
first-order phase transitions, and these can be associated not only to, {\it e.g.}, size-dependent recalescence phenomena in phase change materials~\cite{klimes2020applenergy}, but also to many human diseases that are related to misfolded protein aggregation~\cite{knowles2014natrev}.
	Although several attempts to adapt the classical nucleation theory 
to finite molecular systems have been made ({\it e.g.},~\cite{reguera2003jcp,merikanto2007prl}), most of them 
are based either on the capillarity approximation or on the precary assumption that the shape of 
the aggregates has spherical symmetry.
	However, it may be difficult to imagine that spherically symmetric aggregates are the case when considering, for instance, some biopolymeric systems~\cite{auer2019pnas}. 
	In fact, even for the simplest Ising-like models one may get into trouble when applying the classical theory to systems where molecules present, {\it e.g.}, anisotropic interactions~\cite{cabriolu2012jcp,bingham2013jcp}.

	Recently, the authors of Ref.~\cite{janke2017natcommun} explored the alternative idea that 
free-energy barriers extracted from the microcanonical entropy $S(N,V,E)$ could be used in an 
Arrhenius-like expression, {\it i.e.}, proportional to $e^{-\Delta G^{\dagger}/k_BT}$, to provide 
a shape-free theory for the self-assembly rates in different molecular systems 
({\it i.e.}, Lennard-Jones colloids and polymeric chains).~It
	 could have been a promising approach since equilibrium properties like $\Delta G^{\dagger}$ 
have been already determined by this type of microcanonical shape-free approach in many other 
finite-sized molecular systems which present first-order phase transitions through advanced computational simulations techniques ({\it e.g.}, multicanonical~\cite{berg1992prl,berg2003cpc}, entropic sampling~\cite{lee1993prl}, Wang-Landau~\cite{wanglandau2001prl}, and statistical temperature~\cite{straub2006prl,straub2011jcp,rizzi2011jcp}).~To
	 name a few examples we could mention
polymer adsoption~\cite{wangliang2009jcp,moddel2010pccp} and condensation~\cite{schnabel2011pre}, 
protein folding~\cite{chen2007pre,bereau2010jacs,frigori2013jcp,alves2015cpc}
 and dimerization~\cite{chen2008pre,church2012jcp},
droplet condensation-evaporation~\cite{schierz2016pre}, 
and peptide aggregation~\cite{junghans2006prl,junghans2008jcp,frigori2014pre,trugilho2020jphysconfser}.~Unfortunately, 
	the approach considered in Ref.~\cite{janke2017natcommun}, just like the one 
in Ref.~\cite{frigori2013jcp}, is restricted to the self-assembly kinetics at a temperature
equal to the transition temperature $T^{*}$.
	The full temperature-dependent expressions for the rate constants
was only obtained recently in Ref.~\cite{rizzi2020jstat}, where the proposed
rate theory was able to successfully describe results from both protein 
folding and ice nucleation experiments.








	Here we discuss how one can use 
microcanonical thermostatistics analysis~\cite{schnabel2011pre,qibachmann2018prl} 
as a shape-free and model-independent method to determine the self-assembly rate constants
of macromolecular and colloidal systems that display first-order phase transitions.
	We illustrate our discussion by considering a simple aggregation model for which the
density of states $\Omega(E)$, hence 
 $S(N,V,E)=k_B \ln \Omega(E)$,
can be obtained analytically.

\begin{figure}[!t]
\includegraphics[width=0.38\textwidth]{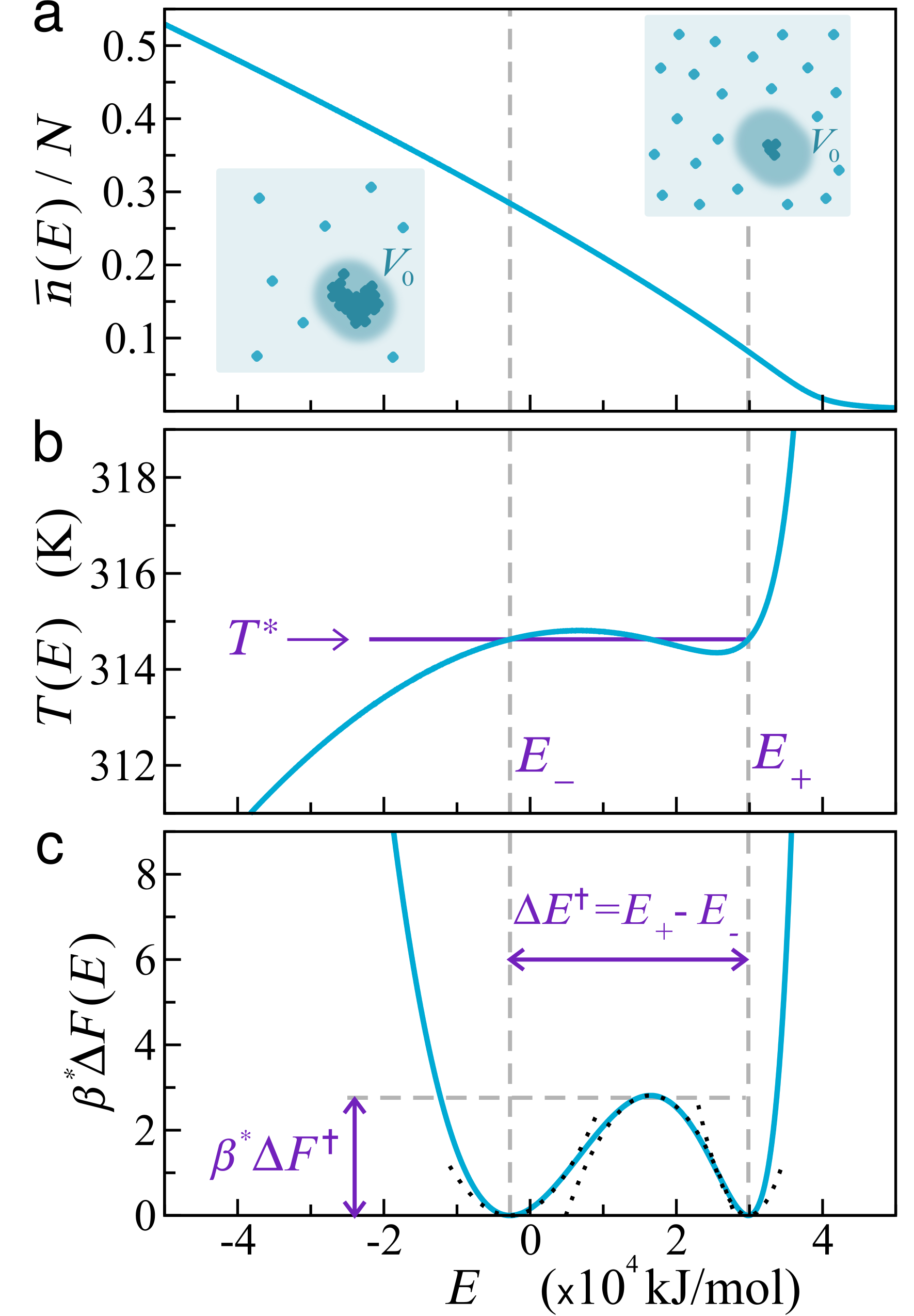}
\caption{(a)~Ratio between the number of particles in the aggregate $\bar{n}(E)$ and the total number of particles $N$,
(b)~microcanonical temperature $T(E)$, and 
(c)~free-energy profile at the transition $\beta^{*}\Delta F(E)$, Eq.~\ref{free-energy-profile}.
Results were obtained for $N=10^4$, $\alpha=1.2$,
$\nu= 3.011\,$kJ/mol, 
and $\eta=7.7$. In (b) the horizontal purple line indicate
the transition temperature $T^{*}=314.626\,$K
determined through a Maxwell-like construction, while the purple 
arrows in (c) denote the free-energy barrier, 
$\beta^{*}\Delta F^{\dagger} = 2.813$, and the microcanonical latent heat, 
$\Delta E^{\dagger} = E_{+}-E_{-} = 3.258 \times 10^{4}\,$kJ/mol.
Dotted lines in (c) are quadratic approximated expressions for $\beta^{*}\Delta F(E)$ (see text).
}
\label{fig:microcanonical}
\end{figure}

	As it is schematically shown in Fig.~\ref{fig:microcanonical}(a),
the model consists of $N$ molecules that are inside a fixed volume $V$,
which is divided into two arbritary volumes, $V_0$ and $V^{\prime}=V-V_0$.
	We assume that the $n$ molecules that are inside the volume
$V_0$ contribute with an interaction energy given by
$E_p(n) = - \nu g(n)$, where $g(n) = \left( n^{\alpha} - 1 \right)$ is
the number of bonds with effective strength $\nu$ between the $n$ molecules.
	The total energy of all molecules in the system is given by $E=E_k+E_p$, 
with $E_k$ being the sum of their kinetic energies. 
	We consider that the molecules in the remaining volume $V^{\prime}$ are diluted 
enough so that they 
do not interact with each other.~Hence, at a given energy $E$, 
the density of states can be evaluated as
\begin{equation}
\Omega(E) \propto \sum_{n=n_{\min}}^{N} \frac{e^{\eta(N-n)}}{n! (N-n)!} (E + \nu g(n))^{3N/2}~~,
\label{omegaE}
\end{equation}
where $\eta = \ln((V - V_0)/V_0)$ and $n_{\min}$ is the minimun number
of molecules that is required to be inside $V_0$ if $E$ is negative 
(note that $E_k$ is always positive and, if $E>0$, then $n_{\min}=1$).
	Because the general ideas that lead to Eq.~\ref{omegaE} are similar to those 
used in Ref.~\cite{campa2016jstat}, a detailed presentation of it (including the phase diagram of the model)
will be discussed elsewhere~\cite{trugilho2021inprep}.
	But, it is worth mentioning that the case $\alpha=2$ corresponds to a 
mean-field long-range interacting model known as the Thirring's model~\cite{latella2015prl,thirring1970zphys}, 
while the case $\alpha=1$ represent a linear-like polymeric system where the molecules interact only 
with their nearest-neighbors. 

	By considering the Stirling's approximation~\cite{chesnut1984amjphys}, and taking the 
number $\bar{n}=\bar{n}(E)$ that maximizes the sum in Eq.~\ref{omegaE} (see Fig.~\ref{fig:microcanonical}(a)),
 the microcanonical entropy can be computed as $S(E) \equiv S(E,\bar{n})=k_B \ln \Omega(E,\bar{n})$.
	Hence, the microcanonical temperature can be evaluated as
$T(E) = 1/k_B b(E)$, where
\begin{equation}
b(E) = \frac{1}{k_B} \left( \frac{\partial S(E)}{\partial E} \right)_{N,V}~~,
\end{equation}
with $k_B$ being the Boltzmann constant.
	Figures~\ref{fig:microcanonical}(a) and~\ref{fig:microcanonical}(b) show, respectively, $\bar{n}(E)$ and $T(E)$ for a system with $N=10^4$ molecules,
$\alpha=1.2$, $\nu=3.011\,$kJ/mol ({\it i.e.},~$5 \times 10^{-21}\,$J per effective bond), 
and $\eta=7.7$. 
	The resulting S-shaped caloric curve obtained for $T(E)$ displayed 
in Fig.~\ref{fig:microcanonical}(b) indicates that the phase transition 
takes place at a temperature $T^{*}=314.626\,$K ({\it i.e.},~$\sim 41.5\,^{\circ}$C).
	The transition temperature separates a high-temperature phase where most of the molecules 
are dissolved in volume $V^{\prime}$, and a low-temperature phase, where 
the system is found at energies closer to $E_{-}$, with 
an aggregate containing $\bar{n}$ molecules formed 
inside the volume $V_0$ (see the schematic drawing in Fig.~\ref{fig:microcanonical}(a)).
	It is important to emphasize that the model does not require any arbitrary information 
about the shape of the volume $V_0$.
	Even so, the value $\alpha<2$ is used here as an effective way to
indicate that some of the molecules must be attached to the periphery of the aggregate.
	In order to obtain experimentally relevant physical units, 
we consider that each molecule have a fixed volume $v_m$,
so that the interacting volume is given by  $V_0=N v_m \approx 4.19 \times 10^{4}\,$nm$^3$, and 
the molar concentration of solute is estimated as $c = N/(N_A V) = \rho/N_A v_m = 179\,\mu$M, 
with $N_A$ being the Avogadro's constant and $\rho = V_{0}/V = (1 + e^{\eta})^{-1} \approx 4.526 \times 10^{-4}$,
{\it i.e.},~$V\approx  9.25 \times 10^{7} \,$nm$^3$.

	At the inverse transition temperature
$\beta^{*} = 1/N_A k_B T^{*}$
 of first-order phase transitions, 
the canonical probability density function (PDF),
$p(E) \propto e^{ - \beta^{*} F(E)}$, present two maxima at energies $E_{-}$ and $E_{+}$, 
and one minimum at $E^{*}$, which correspond to the 
two minima and one maximum in the energy-dependent free-energy $\beta^{*} F(E) =\beta^{*} E - S(E)/k_B$, respectively.
	Thus, the free-energy profile that is shown in Fig.~\ref{fig:microcanonical}(c) can be obtained by considering
\begin{equation}
\beta^{*} \Delta F(E) = 
\beta^{*} F(E) - \beta^{*} F(E_{-})
= ( S^*(E) - S(E) )/k_B~~,
\label{free-energy-profile}
\end{equation}
where $S^*(E) = k_B \beta^{*} (E - E_{-}) + S(E_{-})$,
so that 
$\beta^{*} = k_B^{-1} (S(E_{+}) - S(E_{-}) )/\Delta E^{\dagger}$,
with $S^{*}(E_{+}) \equiv S(E_{+})$, and $\Delta E^{\dagger} = E_{+} - E_{-}$ being the microcanonical latent heat.

	Now, in order to obtain analytical expressions for the forward $\kappa_{-}$ and reverse $\kappa_{+}$ rate constants, we follow the approach discussed in Refs.~\cite{nadler2007pre,szabo2019jcp}, and compute the mean-first passage times (MFPT) $\tau_{-}$ and $\tau_{+}$ directly from estimates for the canonical PDF $p(E)$.
	For instance, the MFPT $\tau_{-}$ which takes for the system to go from the energy $E_{+}$ to $E_{-}$
can be evaluated as
\begin{equation}
\tau_{-} = \frac{1}{D} \int_{E_{-}}^{E_{+}} \frac{dE}{p(E)} \int_{E}^{\infty} p(E') dE'
\approx \frac{4 \pi \tau_{\varepsilon}}{\varepsilon^2 \sqrt{\gamma^{*} \gamma_{+}}} \frac{\Gamma_{+}(\beta)}{\Gamma^{*}(\beta)}~~,
\label{tauminus}
\end{equation}
where $D = \varepsilon^2/2 \tau_{\varepsilon}$ is a diffusion coefficient in the energy space 
that is defined in terms of $\varepsilon$ and $\tau_{\varepsilon}$, which are the typical energy and time scales
involved in the microscopic energy exchange between the $N$ molecules and the thermal reservoir ({\it i.e.}, implicity solvent);
similarly, the MFPT for the system to go from the energy $E_{-}$ to $E_{+}$ is
$\tau_{+} \approx 4 \pi \tau_{\varepsilon} \Gamma_{-}(\beta) /(\varepsilon^2 \sqrt{\gamma^{*} \gamma_{-}}\, \Gamma^{*}(\beta))$,
where the factors $\Gamma^{*}(\beta)$ and $\Gamma_{\pm}(\beta)$ are given, respectively, by 
\begin{eqnarray}
\Gamma^{*}(\beta) &=& \exp \left[ - \beta^{*} \Delta F^{\dagger} + \frac{S^{*}(E_{-})}{k_B} - \beta^{*} E_{-}~~~~~~ \right. \nonumber \\
           &~& \left. ~~~~~ - \frac{[(\beta-\beta^{*}) + \gamma^{*}E^{*}]^2}{2\gamma^{*}} + \frac{\gamma^{*}(E^{*})^2}{2} \right]~,
\label{gammastar}
\end{eqnarray}
and
\begin{eqnarray}
\Gamma_{\pm}(\beta)  &=& \exp \left[\frac{S^{*}(E_{-})}{k_B}  - \beta^{*} E_{-}~~~~~~ \right. \nonumber \\
           &~& \left. 
+ \frac{[(\beta-\beta^{*}) - \gamma_{\pm}E_{\pm}]^2}{2\gamma_{\pm}} - \frac{\gamma_{\pm}(E_{\pm})^2}{2} \right]~.
\label{gammapm}
\end{eqnarray}
	These factors in $p(E)$ can be obtained by writing the microcanonical entropy as $S(E) = S^{*}(E) -k_B \beta^{*} \Delta F(E)$, and expanding the free-energy profile  $\beta^{*} \Delta F(E)$ 
around the energies $E_{\pm}$ as $\beta^{*} \Delta F(E) \approx (\gamma_{\pm}/2)(E - E_{\pm})^2$, and around $E ^{*}$ as $\beta^{*} \Delta F(E) \approx \beta^{*} \Delta F^{\dagger} - (\gamma^{*}/2)(E - E^{*})^2$, with $\beta^{*} \Delta F^{\dagger}$ being a free-energy barrier, as indicated in 
Fig.~\ref{fig:microcanonical}(c).~A 
	numerical fit of these approximated quadratic expressions to the data
displayed in Fig.~\ref{fig:microcanonical}(c) (see dotted lines) yields, {\it e.g.}, 
$\gamma^{*}\approx 4.2 \times 10^{-8}\,$(mol/kJ)$^{-2}$
for $\beta^{*} \Delta F(E)$ close to its maximum at $E^{*}$. 

	By considering the above MFPT 
one can compute the forward 
and reverse 
rate constants, respectively, as
\begin{equation}
\kappa_{-} = \frac{1}{\tau_{-}} \approx A_{+} \exp \left[
-\Delta E_{+}^{\ddagger}(\beta - \beta^{*}) - \frac{\bar{\gamma}_{+}}{2}(\beta - \beta^{*})^2
\right]~~,
\label{kappa_minus}
\end{equation}
and
\begin{equation}
\kappa_{+} = \frac{1}{\tau_{+}} \approx A_{-} \exp \left[
-\Delta E_{-}^{\ddagger}(\beta - \beta^{*}) - \frac{\bar{\gamma}_{-}}{2}(\beta - \beta^{*})^2
\right]~~,
\label{kappa_plus}
\end{equation}
where $\Delta E_{\pm}^{\ddagger} = E^{*} - E_{\pm}$ and $\bar{\gamma}_{\pm} = (\gamma_{\pm})^{-1} + (\gamma^{*})^{-1} $,
with the pre-factors given by
$A_{\pm} = (\varepsilon^2/4\pi\tau_{\varepsilon})\sqrt{\gamma^{*} \gamma_{\pm}}\, e^{-\beta^{*}\Delta F^{\dagger}}$.
	Here it is worth noting that, for the particular case where $\beta=\beta^{*}$,
the above rate constants are indeed proportional to $e^{-\beta^{*}\Delta F^{\dagger}}$,
as empirically suggested in Refs.~\cite{janke2017natcommun,frigori2013jcp}.
	In addition, one can use the above expressions to evaluate the equilibrium rate constant as
\begin{equation}
\kappa_{\text{eq}} = \frac{\kappa_{-}}{\kappa_{+}} \approx A
\exp \left[
\Delta E^{\ddagger}(\beta - \beta^{*}) + \frac{\Delta \bar{\gamma}}{2}(\beta - \beta^{*})^2
\right]~~,
\label{kappa_eq}
\end{equation}
where $A=\sqrt{\gamma_{+}/\gamma_{-}}$, $\Delta E^{\ddagger}=E_{+} - E_{-}$, and
$\Delta \bar{\gamma} = \bar{\gamma}_{-} - \bar{\gamma}_{+}=(\gamma_{-})^{-1} - (\gamma_{+})^{-1}$.




\begin{figure}
\includegraphics[width=0.36\textwidth]{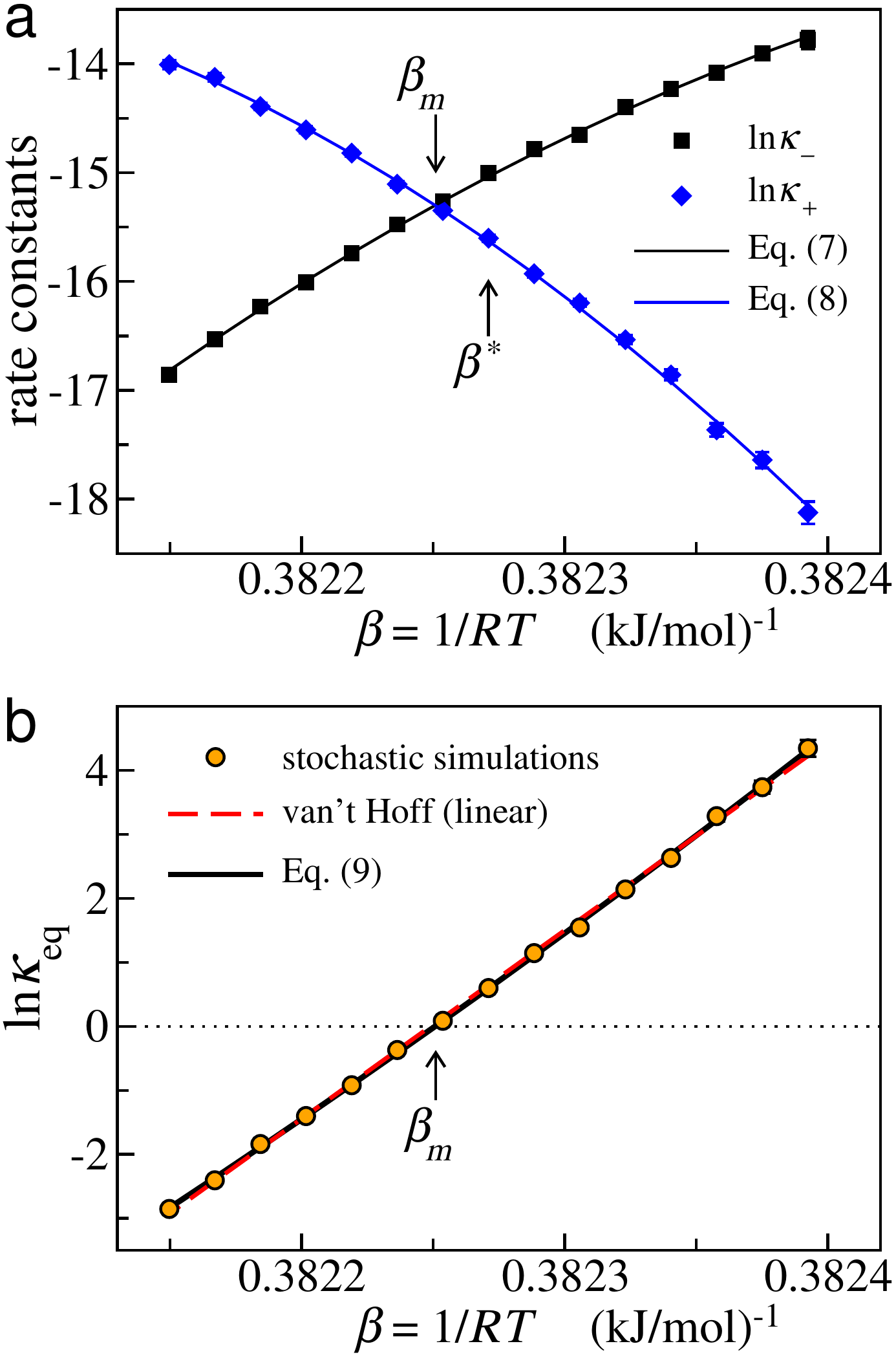}
\caption{(a)~Arrhenius-plot for the forward $\kappa_{-}$ and
reverse $\kappa_{+}$ rate constants.~(b)~Equilibrium rate constant, $\kappa_{\text{eq}} = \kappa_{-}/\kappa_{+}$. 	Arrows indicate the inverse transition temperature 
$\beta^{*}=1/RT^{*}=0.382271\,$(kJ/mol)$^{-1}$,
and the inverse midpoint temperature, 
$\beta_{m}=1/RT_m =0.382251\,$(kJ/mol)$^{-1}$,
where $\kappa_{\text{eq}}=1$.
	Symbols denote numerical results obtained from the stochastic simulations,
while continuous lines correspond to fits to the numerical data.
In (b) we also include the fit using the linear van't Hoff's expression (see text).
}
\label{fig:rates}
\end{figure}



	In order to validate our theoretical approach we have implemented stochastic simulations similar to those described in Ref.~\cite{rizzi2020jstat}, which yield energy time series that lead to stationary distributions given by the canonical PDF, $p(E_j)=  \Omega(E_j)e^{-\beta E_j}/\mathcal{Z}(\beta)$, at discretized energy values, {\it i.e.},
 $E_j= E_{0} + j \varepsilon$ 
with $E_{0}=-\nu (N^{\alpha}-1)$ and 
$\varepsilon=30.11\,$kJ/mol, and
$\mathcal{Z}(\beta)=\sum_{j} \Omega(E_j)e^{-\beta E_j}$ being the canonical partition function~\footnote{The partition function should be evaluated
with the large number summation technique presented in Ref.~\cite{berg2003cpc}.
	In principle, instead of $\Omega(E_j)$ one should use the number of microstates
$\omega(E_j)$ with energies in the range $[E_{j},E_{j+1}[$, but we find no
differences between the results obtained from these two definitions 
for the aggregation model discussed here.}.~In contact
	 with a thermal reservoir at a temperature close to
the transition temperature, the energy fluctuates and the system is able to visit both phases~\cite{rizzi2020jstat}. 
	Hence, one can evaluate the MFPTs $\tau_{-}$ and $\tau_{+}$ numerically by considering the labelled walkers (or coloring/milestoning~\cite{szabo2019jcp}) scheme described in Refs.~\cite{trebst2004pre,nadler2007pre}.



	The numerical results obtained for the rate constants are shown in the Arrhenius plots displayed in Fig.~\ref{fig:rates}.
	The values of the parameters $\alpha$, $N$, $\nu$, and $\eta$ are the same used to produce Fig.~\ref{fig:microcanonical}, which means that the inverse of the transition temperature is equal to
$\beta^{*}=1/RT^{*}=0.382271\,$(kJ/mol)$^{-1}$ (with $R=k_BN_A$). 
	Continuous lines denote the fits to the theoretical expressions, Eqs.~\ref{kappa_minus},~\ref{kappa_plus}, and~\ref{kappa_eq},
from where we got
$\Delta E_{-}^{\ddagger}\approx 1.683 \times 10^{4}\,$kJ/mol, 
$\Delta E_{+}^{\ddagger} \approx -1.264 \times 10^{4}\,$kJ/mol,
and 
$\Delta E^{\ddagger} \approx 2.947\times 10^{4}\,$kJ/mol,
	by assuming that the barrier height is given by the microcanonical estimate, 
{\it i.e.},~$\beta^{*} \Delta F^{\dagger} = 2.813$, and that 
$\gamma^{*} \approx 4.2 \times 10^{-8}\,$(kJ/mol)$^{-2}$. 
	This fitting procedure also yields
$\gamma_{-} \approx 3.4 \times 10^{-8}\,$(kJ/mol)$^{-2}$
and 
$\gamma_{+} \approx 11 \times 10^{-8}\,$(kJ/mol)$^{-2}$,
so that the values of $\bar{\gamma}_{-}$ and $\bar{\gamma}_{+}$ are self-consistent 
with the pre-factores $A_{-}$, $A_{+}$, and $A$.
	Importantly, the obtained values for $\gamma_-$ and $\gamma_+$ are also in 
good agreement with the values one would obtain 
by considering direct fits of the quadratic approximated expressions to the free-energy profile
$\beta^{*}\Delta F(E)$ at the minima displayed in Fig.~\ref{fig:microcanonical}(c)
(in fact, these values were used to plot the dotted curves close to $E_{\pm}$ in that figure).

	As noted in Ref.~\cite{rizzi2020jstat}, the pre-factors $A_{-}$ and $A_{+}$ may be different 
if the wells of the free-energy profile are asymmetrical, so that the equilibrium rate $\kappa_{\text{eq}}$ is 
not necessarily equal to one at the transition temperature $T^{*}$.
	Indeed, the asymmetry observed in the wells of the free-energy profiles $\beta^{*}\Delta F(E)$ displayed in Fig.~\ref{fig:microcanonical}(c) is consistent with the pre-factor 
$A=1.805$
 and the positive value found for 
$\Delta \bar{\gamma} \approx  0.205 \times 10^{8}\,$(kJ/mol)$^{2}$, 
which means that the well close to the energy $E_{+}$ is sharper than the well close 
to $E_{-}$ ({\it i.e.}, $\gamma_{-}<\gamma_{+}$).
	The midpoint transition temperature 
$T_{m}=1/R \beta_{m}$
can be estimated from Eq.~\ref{kappa_eq} by considering that 
$\beta_{m} = \beta^{*} + \delta$ and imposing that 
 $\kappa_{\text{eq}}(T_m) = 1$, which yields $\delta = (2 \Delta E^{\ddagger})^{-1} \ln(\gamma_{-}/\gamma_{+})$.
	Hence, 
at 
$T_{m}=314.642\,$K, one finds that the peaks in $p(E)$ should have
the same area under the curve, instead of having maxima with equal heights as at $T^*$.

	It is worth noting that, although the logarithm of the equilibrium rate in Fig.~\ref{fig:rates}(b) seems to display a linear behavior that can be fitted to the usual van't Hoff expression, {\it i.e.}, $\kappa_{\text{eq}}^{\ominus} = \exp [ \beta \Delta E^{\ominus} - \Delta S^{\ominus}/k_B ]$, the forward and reverse rates in Fig.~\ref{fig:rates}(a) clearly show non-Arrhenius behaviors.
	The linear fit to $\ln \kappa_{\text{eq}}$ yields
$\Delta E^{\ominus}=2.947 \times 10^4\,$kJ/mol
and 
$\Delta S^{\ominus}/k_B=1.126 \times 10^4$, 
which are close to the values $\Delta E^{\ddagger}$ and $\Delta S^{\ddagger}/k_B =  \beta^{*}\Delta E^{\ddagger}$ obtained from the fit of Eq.~\ref{kappa_eq} to the numerical data. 
	Accordingly, as it is noted in Ref.~\cite{rizzi2020jstat}, the values of 
$\Delta E^{\ddagger}$ and $\Delta S^{\ddagger}$ that are obtained from the kinetic approach are 
systematically close to the equilibrium values extracted from 
Fig.~\ref{fig:microcanonical}(c), {\it i.e.}, $\Delta E^{\ddagger} \approx 0.9 \Delta E^{\dagger}$
and $\Delta S^{\ddagger} \approx 0.9 \Delta S^{\dagger}$, even though
we assume that the energies $E_{-}$ and $E_{+}$ 
are independent of the temperature.
	Importantly, the latent heat obtained from the
kinetic approach is consistent with 
the value of the effective interaction energy per bond, 
{\it i.e.}, $\Delta E^{\ddagger}/(N N_A) \sim 5 \times 10^{-21}\,$J.


\begin{figure}[!t]
\includegraphics[width=0.48\textwidth]{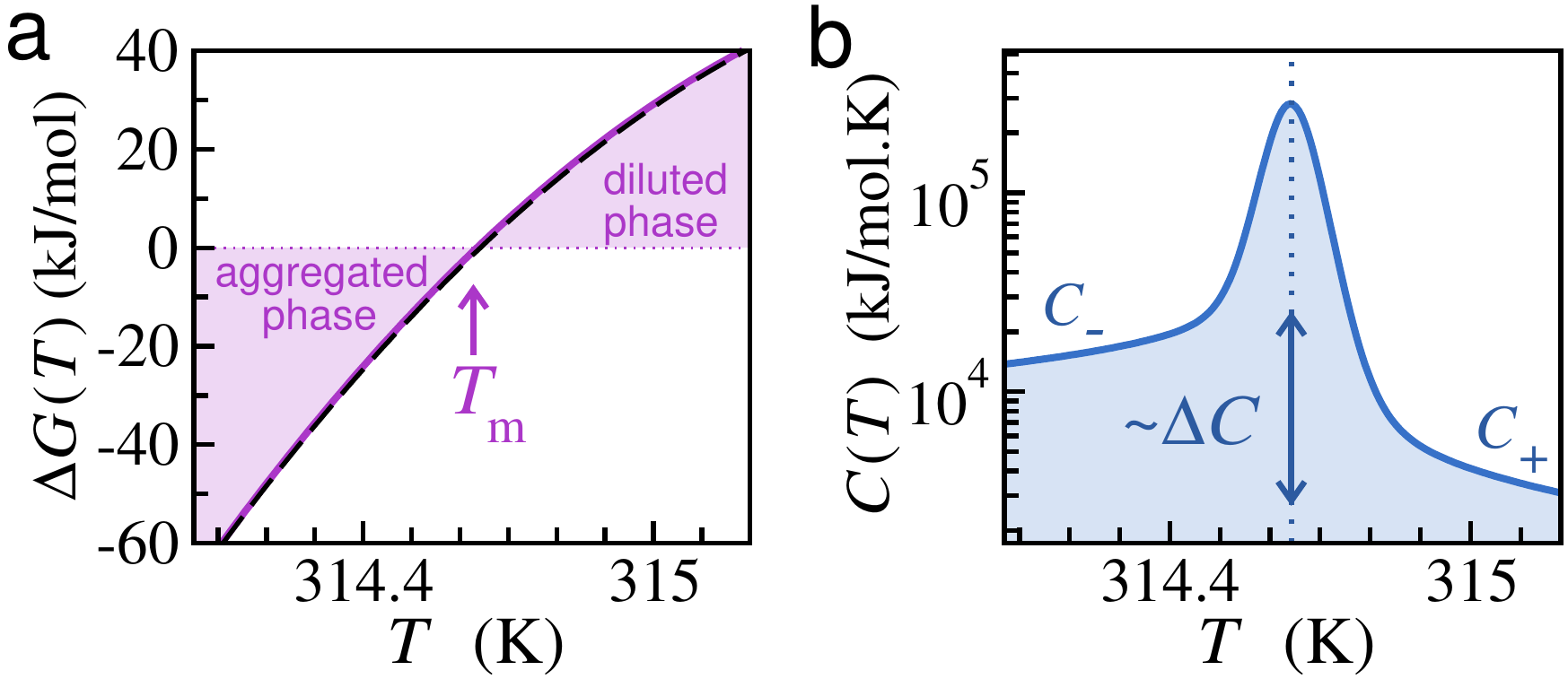}
\caption{(a)~Effective free-energy $\Delta G(T)$, Eq.~\ref{deltaGT}, 
obtained with the parameters extracted from Fig.~\ref{fig:rates}, and its numerical fit
to Eq.~\ref{deltaGTmodel} (dashed black line).
(b)~Heat capacity $C(T)$ close to the transition temperature $T_{m}$ (dotted line),
with $\Delta C <0$, obtained from the PDF evaluated from the density of states, Eq.~\ref{omegaE},
with the parameters used in Fig.~\ref{fig:microcanonical}.
}
\label{fig:effectiveG}
\end{figure}


	Finally, we include in Fig.~\ref{fig:effectiveG}
the temperature-dependent effective free-energy, 
$\Delta G(T) = - RT \ln \kappa_{\text{eq}}$, 
and the 
heat capacity, 
$C(T)=  \left( \langle E^2 \rangle - \langle E \rangle^2 \right)/RT^2$,
in order to illustrate the relationship between our theory and
the thermal analyses that are commonly used to interpret the experimental data~\cite{hohne2003DSC}.
	From Eq.~\ref{kappa_eq}, we find 
\begin{equation}
\Delta G(T) \approx - \Delta E^{\ddagger} \left( 1 - \frac{T}{T_m} \right)
- \frac{\Delta \bar{\gamma}}{2 RT} \left( 1 - \frac{T}{T_{m}} \right)^2~,
\label{deltaGT}
\end{equation}
which can be compared, for instance, with a widely used expression for macromolecular systems
given by
\begin{equation}
\Delta G(T) \approx - \Delta E^{\ddagger}\left( 1 - \frac{T}{T_m} \right)
- \Delta C \left[ (T - T_{m})- T \ln \left( \frac{T}{T_{m}} \right) \right],
\label{deltaGTmodel}
\end{equation}
that is obtained from the assumption that $\Delta C$ is an effective temperature-independent
change in the heat capacity~\cite{baldwin1986pnas}.
	Although the latter approximation (Eq.~\ref{deltaGTmodel}) might be poor in some cases, the fit of Eq.~\ref{deltaGTmodel} to the data displayed in Fig.~\ref{fig:effectiveG}(a) yields
 $\Delta C \approx -2.3 \times 10^{4}\,$kJ/mol.K,
which can be used to verify that~\cite{rizzi2020jstat}
$\Delta C \approx - \Delta \bar{\gamma}/RT_{m}^{2}< 0$ when $T \approx T_m$.
	Indeed, the behavior of $C(T)$ displayed in Fig.~\ref{fig:effectiveG}(b)
for the aggregation transition studied here, {\it i.e.}, with $C_{-} > C_{+}$ and $\Delta C<0$,
is also supported by numerical simulations of more detailed aggregation models~\cite{irback2013prl,irback2015jcp,nussbaumer2016jphysconfser,janke2017jphysconfser}.
	Even so, 
our model-independent kinetic approach can be also applied to study the 
behavior observed for, {\it e.g.}, protein folding transitions~\cite{cooper2010jphyschemlett}, where
one may find that~\cite{rizzi2020jstat} $\Delta C>0$ and $\Delta \bar{\gamma}<0$.

	In conclusion, we present a shape-free rate theory that
allows one to associate equilibrium properties 
that are determined from an analysis based on the microcanonical entropy
$S(N,V,E)$ to the self-assembly kinetics of finite-sized macromolecular 
and colloidal systems.
	It is worth mentioning that our approach provides an 
insightful theoretical interpretation for the temperature-dependent 
rate constants, {\it i.e.}, Eqs.~\ref{kappa_minus},~\ref{kappa_plus}, and~\ref{kappa_eq}, 
in contrast to the popularly used Arrhenius-like expressions that are based on phenomenological 
approaches~\cite{laidler1984jchemedu,vanthoffbook}.
	Hence, we believe that our theory should help experimentalists that work with 
phase change materials~\cite{klimes2020applenergy} and nucleation phenomena~\cite{sear2007jphys,vekilov2016progcryst}
to interpret their results. 
	In particular, one could use the 
kinetic approach presented here as a reliable method to reconstruct 
free-energy profiles $\beta^{*}\Delta F(E)$ and microcanonical entropies $S(E)$
from the experimental kinetic data.




	The  authors acknowledge the funding by the Brazilian agencies CAPES (code 001), FAPEMIG 
(Process\,APQ-02783-18), and CNPq (Grants 306302/2018-7 and 426570/2018-9).

\vspace{-0.5cm}



%

\end{document}